\documentclass[12pt]{iopart}
\usepackage{iopams}  
\usepackage{graphicx}
\usepackage{color}

\begin{document}


\newcommand{\be}{\begin{eqnarray}}
\newcommand{\ee}{\end{eqnarray}}
\newcommand{\mathbfe}{\begin{subequations}}
\newcommand{\ese}{\end{subequations}}


\newcommand{\bnum}{\begin{enumerate}}
\newcommand{\enum}{\end{enumerate}}

\newcommand{\bit}{\begin{itemize}}
\newcommand{\eit}{\end{itemize}}

\newcommand{\bc}{\begin{cases}}
\newcommand{\ec}{\end{cases}}


\newcommand{\erf}{\mrm{erf}}


\newcommand{\bpm}{\begin{pmatrix}}
\newcommand{\epm}{\end{pmatrix}}

\newcommand{\bvm}{\begin{vmatrix}}
\newcommand{\evm}{\end{vmatrix}}


\newcommand{\mbf}{\mathbf}
\newcommand{\mbb}{\mathbb}
\newcommand{\mcal}{\mathcal}
\newcommand{\mfr}{\mathfrak}
\newcommand{\mrm}{\mathrm}

\newcommand{\ul}{\underline}
\newcommand{\ovl}{\overline}

\newcommand{\vol}{\mrm{vol}}



\newcommand{\ga}{\alpha}
\newcommand{\gb}{\beta}
\newcommand{\gc}{\gamma}
\newcommand{\gd}{\delta}
\newcommand{\eps}{\epsilon}
\newcommand{\gf}{\phi}
\newcommand{\gl}{\lambda}
\newcommand{\gk}{\kappa}
\newcommand{\go}{\omega}
\newcommand{\gt}{\theta}
\newcommand{\gr}{\rho}
\newcommand{\gs}{\sigma}

\newcommand{\Gf}{\Phi}
\newcommand{\Go}{\Omega}
\newcommand{\Gc}{\Gamma}
\newcommand{\Gt}{\Theta}
\newcommand{\Gd}{\Delta}
\newcommand{\Gs}{\Sigma}
\newcommand{\Gl}{\Lambda}

\newcommand{\gve}{\epsilon}
\newcommand{\gvf}{\varphi}
\newcommand{\gvr}{\varrho}

\newcommand{\h}{\hbar}

\newcommand{\p}{\partial}
\newcommand{\f}{\frac}
\newcommand{\diff}{\mrm{d}}
\newcommand{\iy}{\infty}
\newcommand{\lap}{\triangle}
\newcommand{\nab}{\nabla}

\title[Minimal  continuum models  of  active fluids]{Minimal continuum theories of structure formation in dense active fluids}


\author{J\"orn Dunkel\dag
\footnote[3]{To whom correspondence should be addressed (jd548@cam.ac.uk)}\ , 
Sebastian Heidenreich\ddag\ ,  
Markus B\"ar\ddag\  
and 
Raymond E. Goldstein\dag 
}

\address{\dag\ DAMTP, Centre for
Mathematical Sciences, University of Cambridge, Wilberforce Road, Cambridge CB3 0WA, UK}

\address{\ddag\ Physikalisch-Technische Bundesanstalt,
Abbestr. 2-12, 10587 Berlin, Germany}

\begin{abstract}
Self-sustained dynamical phases of living matter can exhibit remarkable similarities over a wide range of scales, from  mesoscopic vortex structures in  microbial suspensions  and motility assays of biopolymers to turbulent large-scale instabilities in flocks of birds or schools of fish. Here, we argue that, in many cases, the phenomenology of such active states can be efficiently described in terms of  fourth- and higher-order partial differential equations. Structural transitions in these models  can be interpreted as Landau-type kinematic transitions in Fourier (wavenumber) space, suggesting that microscopically different biological systems can share universal long-wavelength features. This general idea is illustrated through numerical simulations for two classes of continuum models for incompressible active fluids: a Swift-Hohenberg-type scalar field theory, and  a minimal vector model that extends the classical Toner-Tu theory and appears to be a promising candidate for the quantitive description of dense bacterial suspensions. We also discuss briefly how microscopic symmetry-breaking mechanisms can enter macroscopic continuum descriptions of collective microbial motion near surfaces\color{black}{, and  conclude by outlining future applications.}
\end{abstract}


\submitto{\NJP}

\maketitle

\section{Introduction}

Simple and complex life forms can exhibit remarkably similar collective behaviors over a   wide range of length and time scales~\cite{2012Vicsek,2010Ramaswamy,2011KochSub}. Well-known examples are  flocking phenomena in swarms of birds~\cite{2009Parisi} and self-sustained turbulent phases in  schools of fish~\cite{2011Couzin} that share  several qualitative features with the meso-scale dynamics in bacterial suspensions~\cite{1997Kessler,2004DoEtAl,2007Cisneros,2012Lin} and  films~\cite{2007SoEtAl,Swinney_bactclust,2012Peruani}. 
{\color{black}{When studying such processes from a physicist's perspective, a main challenge consists in identifying generic  models that capture the most essential aspects of their dynamics.  To this end, it is often useful to regard biological systems  that comprise  a large number of elementary self-propelled units, such as motor proteins or swimming cells in suspensions, as active~\lq fluids\rq~\cite{2010Ramaswamy,2011KochSub, 2012Marchetti,2012Wensink}. Unlike conventional liquids, which typically require external energy injection (stirring, shearing, shaking, etc.) for the formation of large-scale patterns, active fluids are driven internally as their microscopic constituents are capable of transforming chemical into kinetic energy.  The interplay between this  intrinsic pumping and nonlinear elastic stresses due to physical or biological interactions facilitates the emergence of complex dynamical structures~\cite{2012Vicsek,2010Ramaswamy,2011KochSub}, whose systematic classification poses a formidable theoretical task.}
}

\par
Over the past two decades,  intense efforts have been made to understand the phenomenology of microbial and other active fluids, but in spite of substantial progress it is still not entirely clear which of their characteristics are universal or system-specific~\cite{2010Ramaswamy,2012Toner,2012PawelReview}, and which classes of dynamical equations are capable of providing adequate minimal descriptions. In recent years,  a considerable number of continuum models for active systems have been proposed~\cite{2010Ramaswamy,2005ToTuRa,1998TonerTu_PRE,2008Wolgemuth,2009BaMa_PNAS,LB_Marenduzzo_hybrid,2010Pedley,2002Ra,2008SaintillanShelley}, but most of them have yet to be tested against experiments~\cite{2012Wensink,2007Ar}.  Many of those theories focus on the couplings between two or more order-parameters (concentration, solvent velocity, orientation fields, etc.) and typically involve a large number of parameters, thus making comparison with experimental data very difficult. 
To exploit current and future progress in experimental imaging and  tracking techniques~\cite{2011Couzin,2010Bausch,2011Japan,2012Lu,2012Sumino}, and to understand better the general ordering principles that govern active matter~\cite{2012Vicsek,2010Ramaswamy}, it will be necessary to identify tractable minimal models that not only capture the essential instability mechanisms but also allow for quantitative comparison with experiments. 

\par 

In this paper, we will analyze two such minimal continuum theories for active suspensions by focussing  
on generic  structural properties and stressing formal analogies with classical phase transitions. 
Our approach is based on the hypothesis that dynamical transitions in many internally or externally driven systems, such as microbial~\cite{1997Kessler,2004DoEtAl,2007Cisneros,2012Wensink,2012Lu,2005Riedel_Science}  or vibrated colloidal suspensions~\cite{2006Aranson,2009Herminghaus}, can be phenomenologically modeled as Landau-type transitions in Fourier (wave-number) space, which suggests that minimal hydrodynamic descriptions  of active matter can be obtained in terms of higher-than-second-order partial differential equations (PDEs).  Higher-order PDEs have been previously derived and studied for a wide range of nonlinear structure formation phenomena~\cite{2006Aranson,1990Knobloch,1991Ray,2009Cross,2010Nicoli}, including Rayleigh-Benard convection~\cite{1977SwiftHohenberg}, polymer and vesicle dynamics~\cite{2006Du},  quasi-crystal formation~\cite{1997Lifshitz} and theories of ionic liquids~\cite{2011Bazant}. However, to our knowledge, models of this type have rarely been considered in the context of microbial suspensions~\cite{2011Boyer}. Therefore, one of our main objectives here is to draw attention to the possibility that  one can obtain useful, testable continuum theories of bacterial and other active fluids, by restricting field variables to a minimal set of experimentally accessible  order-parameters but admitting fourth- and higher-order spatial derivatives. {\color{black}{Such theories can then be used as phenomenological models to obtain predictions for the behavior of active fluids under shear~\cite{2012Marenduzzo} or in different confining geometries~\cite{2012Woodhouse}, thereby providing a conceptual basis for the interpretation of rheological measurements~\cite{2009Rafai,2009SoAr,2011Aranson_PRE} and the  design of optimized microfluidic devices for the control of microbial flow~\cite{2010Austin_PRL,2012Kantsler_PNAS}.
}}
\par

The basic idea is readily summarized as follows:
It is well-known that the incompressible Navier-Stokes equation is capable of describing the dissipative flow dynamics $\mathbf{v}(t,\mathbf{x})$ of a wide range of conventional \lq passive\rq\space fluids, regardless of their exact microscopic composition. A main reason for this is that these systems behave similarly at long-wavelengths (small-wave numbers), so that the leading order viscous dissipation can be described by a term  $\Gamma_0\triangle \mathbf v$ in the field equations, with partial information about the microstructure being retained in the viscosity coefficient $\Gamma_0$ (throughout, $\triangle =\nabla^2$ denotes the Laplacian). Transforming to Fourier-space, the dissipative term  yields a simple quadratic \lq dispersion\rq\space relation $\sim\Gamma_0 |\mathbf k|^2$, which represents the dominant contribution in a systematic small-wavenumber expansion and leads to damping in the absence of external stimuli.  
By contrast, in active fluids, viscous dissipation competes with internal or external energy input and, in principle, one cannot exclude that higher-order contributions  of the form $\Gamma_0 |\mathbf k|^2+\Gamma_2 |\mathbf k|^4+\ldots$ become relevant as well.  In fact, they will certainly be needed to ensure stability if, due to the complex interplay of nonlinear interactions and energy input, the coefficient $\Gamma_0$ should change its sign. 
{\color{black}{In this context, it should be noted that, in the case of active fluids, the coefficients $\Gamma_n$ and other transport coefficients~\cite{2008BaMa,2009Bertin,2012Peshkov_PRL} will depend on both the physical interaction parameters and the motility parameters of the microscopic fluid constituents.}  
}
Formally, the inclusion of higher-order terms in the Fourier-space expansions is analogous to the well-known Landau-expansion of order-parameter potentials and, accordingly,  sign-changes in the coefficients $\Gamma_n$ can give rise to Landau-type kinematic phase transitions.  When going back to position space, terms $|\mathbf k|^4, |\mathbf k|^6,\ldots $ will transform into higher-order spatial derivatives $\triangle^2, \triangle^3,\ldots$. The inclusion of such terms\footnote{The restriction to functions of $|\mathbf k|$ is dictated by isotropy; in principle, one could also study odd and fractional powers of $|\mathbf k|$ but this would go  beyond the scope of the present paper. 
} makes the theory successively more non-local.  {\color{black}{The physical origins of effectively  non-local interactions can be manifold~\cite{1981Volkenstein,1981Volkenstein_2,2007Hutt}, ranging from global packing constraints to hydrodynamic and chemical interactions in biological systems. In microbial suspensions, such non-localities may arise naturally from active stresses that are generated by the swimming strokes of the organisms and transported through the fluid. It seems plausible that non-local stress contributions are also present in passive fluids, even though they are not dynamically relevant in this case since friction is dominated by the Laplacian viscosity term. By contrast, for active systems, recent studies~\cite{2012Wensink} have shown that a model with a negative coefficient $\Gamma_0$  captures experimental observation like energy spectra and correlation functions in a quantitative manner.  Such models need to include the higher-order derivative terms  as they provide the necessary damping at small wave-lengths.
} 
}

\par

From the preceding considerations, it seems plausible that a systematic characterization of active fluids in terms of their asymptotic small  wave-number expansions can help to distinguish specific from universal properties, thereby providing a basis for more systematic classification schemes similar to those for  thermodynamic equilibrium phases in classical fluids or spin systems. Moreover, this analogy-driven approach promises analytically tractable models of active suspensions that are considerably simpler than many of the currently studied  (potentially more accurate) multi-component theories~\cite{2008Wolgemuth,2009BaMa_PNAS,LB_Marenduzzo_hybrid}, and will hopefully enable quantitative comparisons with experiments in the near future. In the present paper, we shall focus on theoretical aspects of fourth-order continuum models, starting with the simplest case, which is given by a Swift-Hohenberg-type scalar or pseudo-scalar field theory~\cite{1990Knobloch,1977SwiftHohenberg}. This model is used as a basic example to illustrate how microscopic symmetry-breaking mechanisms~\cite{2005Berg_Nature} can enter macroscopic continuum descriptions of microbial motion near surfaces~\cite{2005Riedel_Science}.  
Subsequently, we will generalize to  non-scalar order-parameters by considering a minimal vector theory for incompressible active suspensions. The resulting flow model extends the seminal Toner-Tu theory~\cite{2012Toner,1998TonerTu_PRE} and is a promising candidate for the quantitive description of highly concentrated bacterial fluids~\cite{2012Wensink}.  In the subsequent  discussion, we use  results from two-dimensional (2D) continuum simulations to illustrate selected dynamical properties of the different models in more detail.

\section{(Pseudo) scalar order-parameter theory}
\label{sec:scalar}

The minimal model considered in this section belongs to the class of generalized Swift-Hohenberg theories~\cite{2006Aranson,1977SwiftHohenberg}. Our motivation for prepending a brief discussion of this well-known model here is two-fold: It is helpful to recall some of its basic properties before considering the generalization to vectorial order-parameters. This model is also useful for illustrating how microscopic symmetry-breaking mechanisms~\cite{2005Berg_Nature} can be incorporated into macroscopic descriptions of experimentally relevant microbial systems~\cite{2005Riedel_Science}, as discussed in Section~\ref{s:symmetry_breaking} below.

\subsection{Model equations}

We consider the simplest  isotropic fourth-order model for a non-conserved scalar or pseudo-scalar order-parameter~$\psi(t,\mathbf x)$, given  by
\begin{equation}\label{e:scalar}
 \partial_t \psi = F(\psi)+\gamma_0 \Delta \psi - \gamma_2 \Delta^2 \psi ,
\end{equation}
where $\p_t=\p/\p t$ denotes the time derivative, and $\triangle=\nabla^2$ is the $d$-dimensional Laplacian. The force $F$ is derived from a Landau-potental $U(\psi)$
 \begin{equation}\label{e:scalar_potential}
F=-\frac{\partial U}{\partial \psi},
\qquad\qquad  
U(\psi) = \f{a}{2}\psi^2 + \f{b}{3} \psi^3 + \f{c}{4} \psi^4.
 \end{equation}
We will assume throughout that the system is confined to a finite spatial domain $\Go\subset \mathbb{R}^d$ of volume
\be
|\Go|=\int_\Omega d^dx,
\ee
adopting with periodic boundary conditions in simulations. The derivative terms on the rhs.~of~\Eref{e:scalar} can also be obtained by variational methods from a suitably defined energy functional (see \ref{app:scalar}). In the context of active suspensions, $\psi$ could, for example, quantify local energy fluctuations,  local alignment, phase differences, or vorticity.  
{\color{black}
In this case, the transport coefficients  $(a,b,c,\gc_1,\gc_2)$ in Equations~\eref{e:scalar} and \eref{e:scalar_potential} will contain passive contributions due to steric or other physical interactions  as well as active motility-related  contributions. In general, it is very challenging to derive the exact functional dependence between macroscopic transport coefficients and microscopic interaction and motility parameters for active non-equilibrium systems~\cite{2008BaMa,2009Bertin,2012Peshkov_PRL,2012Grossmann}. With regard to practical applications, however, it is often sufficient to view transport coefficients as purely phenomenological  parameters that can be determined by matching the solutions of continuum models, such as the one defined by Equations~\eref{e:scalar} and \eref{e:scalar_potential}, to experimental data~\cite{2012Wensink}. This is analogous to treating the viscosity in the classical Navier-Stokes equations as a phenomenological fit parameter. The actual predictive strength of a continuum model lies in the fact that, once the parameter  values have been determined for given a set-up, the theory can be used to obtain predictions for how the system should behave in different geometries or under changes of the boundary conditions (externally imposed shear, etc.). 
In some cases, it may also be possible to deduce qualitative parameter dependencies from physical or biological considerations. For instance, if $\psi$ describes the vorticity of an isolated active fluid, say a bacterial suspension, then transitions from $a>0$ to $a<0$ or $\gc_0>0$ to $\gc_0<0$, which both lead to non-zero flow patterns, must be connected to the microscopic self-swimming speed $v_0$ of the bacteria. Assuming a linear relation, this suggests that, to leading order, $a_0=\delta- \alpha v_0 $  where $\delta >0$ is a passive damping contribution and $\alpha v_0>0$ the active part, and similarly for $\gc_0$.
}
 \par
For completeness, one should also note that in the case of a conserved order-parameter field $\varrho$ the field equations would either have to take the current-form $\partial_t \varrho=-\nabla\cdot {\mathbf J}(\varrho)$ or, alternatively, 
one could implement conservation laws globally by means of Lagrange  multipliers~\cite{2006Du}. For example,  for a dynamics similar to that of \Eref{e:scalar} and a simple global \lq mass\rq\space constraint
\be\nonumber
M=\int_\Omega d^dx\, \varrho = const,
\ee
the Lagrange-multiplier  approach yields the non-local  equations of motions
 \begin{eqnarray}
 \nonumber
  \partial_t \varrho &= F(\varrho) +\gamma_0 \Delta \varrho - \gamma_2 \Delta^2 \varrho -\gl_1, \\
  \gl_1&=
  \f{1}{|\Omega|}\int_\Omega d^dx\,\left[ F(\varrho)+\gamma_0 \Delta \varrho - \gamma_2 \Delta^2 \varrho \right].
  \nonumber
 \end{eqnarray}
 In the remainder of this section, however, we shall focus on the local dynamics defined by Equations~\eref{e:scalar}
 and~\eref{e:scalar_potential}, since this well-known example will be a useful reference point for the discussion of the vector model in Section~\ref{sec:vector}.
 
\subsection{Linear stability}

The fixed points of \Eref{e:scalar} are determined by the zeros of the force $F(\psi)$, corresponding to the minima of the potential $U$, yielding $\psi_0=0$ and 
\be
\psi_\pm=-\f{b}{2c} \pm \sqrt{\f{b^2}{4c^2} -\f{a}{c}},
\qquad
\mathrm{if}\quad b^2>4ac.
\ee 
Linearization of \Eref{e:scalar} near $\psi_0$ for small perturbations $\psi=\eps_0 \exp(-\gs_0 t -i\mathbf k\cdot \mathbf x)$ gives
\be\label{e:psi_0}
\gs_0(\mathbf k)=a+\gc_0 |\mathbf k|^2+\gc_2|\mathbf k|^4.
\ee
Similarly, one finds for $\psi =\psi_\pm+\eps_\pm \exp(-\gs_\pm t -i\mathbf k\cdot \mathbf x)$
\be\label{e:psi_pm}
\gs_\pm(\mathbf k)= -(2a+b\psi_\pm)+\gc_0 |\mathbf k|^2+\gc_2|\mathbf k|^4.
\ee
The unusual sign-convention in the exponential of the perturbation ansatz was so chosen as to emphasize the formal similarity of Equations~\eref{e:psi_0} and \eref{e:psi_pm} with the quartic Landau potential~\eref{e:scalar_potential}, i.e., modes with $\gs<0$ are unstable.

From Equations~\eref{e:psi_0} and~\eref{e:psi_pm}, we see immediately that $\gc_2>0$ is required to ensure small-wavelength stability of the theory and, furthermore, that non-trivial dynamics can be expected if $a$ and/or $\gc_0$ take  negative values. In particular, all three fixed points can become simultaneously unstable if $\gc_0<0$. The analogy with classical Landau-transitions is evident if we compare~\eref{e:psi_0} and~\eref{e:psi_pm} with the order-parameter potential $U$ in \Eref{e:scalar_potential} for the  symmetric case $b=0$: Changing the sign of $\gc_0$ induces a dynamical transition (in Fourier space), which is formally similar to the standard \lq configurational\rq\space  second-order transition~\cite{2010Ramaswamy} in the vicinity of $a=0$.

\subsection{Numerical results in 2D}
\label{s:scalar_numerics}
We briefly illustrate the $\gc_0$-induced changes in the dynamics of the (pseudo-)scalar field~$\psi(t,\mathbf x)$ through 2D numerical results. The discussion in this part merely serves as a reminder before considering  symmetry-breaking  in Section~\ref{s:symmetry_breaking}.

\paragraph{Algorithm}
To simulate Equations~\eref{e:scalar}
 and~\eref{e:scalar_potential} in two space dimensions, we implemented a pseudospectral algorithm with periodic boundary conditions as commonly used in computational fluid dynamics \cite{Orszag}. The model equations are projected onto a Fourier space basis, and the remaining ordinary differential equations are solved numerically by an operator splitting method that computes the linear operator exactly \cite{Pedrosa}. The nonlinear terms were evaluated by applying the \lq 2/3-rule\rq~to suppress aliasing errors  \cite{Canuto}.   We simulated the model dynamics on 2D cubic grids with sizes ranging from $64\times 64$ to $256\times 256$ lattice points. The solver was written in Matlab, and its numerical stability was verified for a wide range of parameters and space-time discretizations. Rescaled dimensionless variables and parameters as adopted in the simulations are summarized in Table~\ref{para_psi}. The rescaled time steps  were typically of the order of  $\Delta t = 10^{-1}$.  All simulations were initiated with isotropic, randomly chosen order-parameter values.

\begin{table}[t]
\centering
\begin{tabular}{c|c}
model parameter & rescaled dimensionless parameter \\
\hline
$a$ &  $ a \;t_\mathrm{u}$\\
$b$ & $b \;\psi_\mathrm{u} \; t_\mathrm{u} $\\
$c$ & $1$ \\
\hline
$\gamma_0$ & $  \gamma_0\;t_\mathrm{u}L^{-2}$ \\
$\gamma_2$ & $1$
\end{tabular}
\caption{The right column shows the rescaled dimensionless parameters used in the simulations of the (pseudo)scalar model from Equations~\eref{e:scalar}
 and~\eref{e:scalar_potential}. The unit time is defined by the damping time-scale $t_\mathrm{u} =L^4 /\gamma_2$, where $L$ is the length of the 2D simulation box,  and the order-parameter is measured in units of  $\psi_\mathrm{u} =1/\sqrt{t_\mathrm{u}c}$.
 \label{para_psi}}
\end{table}

\begin{figure} [t]
\centering
  \includegraphics[width=13.0cm]{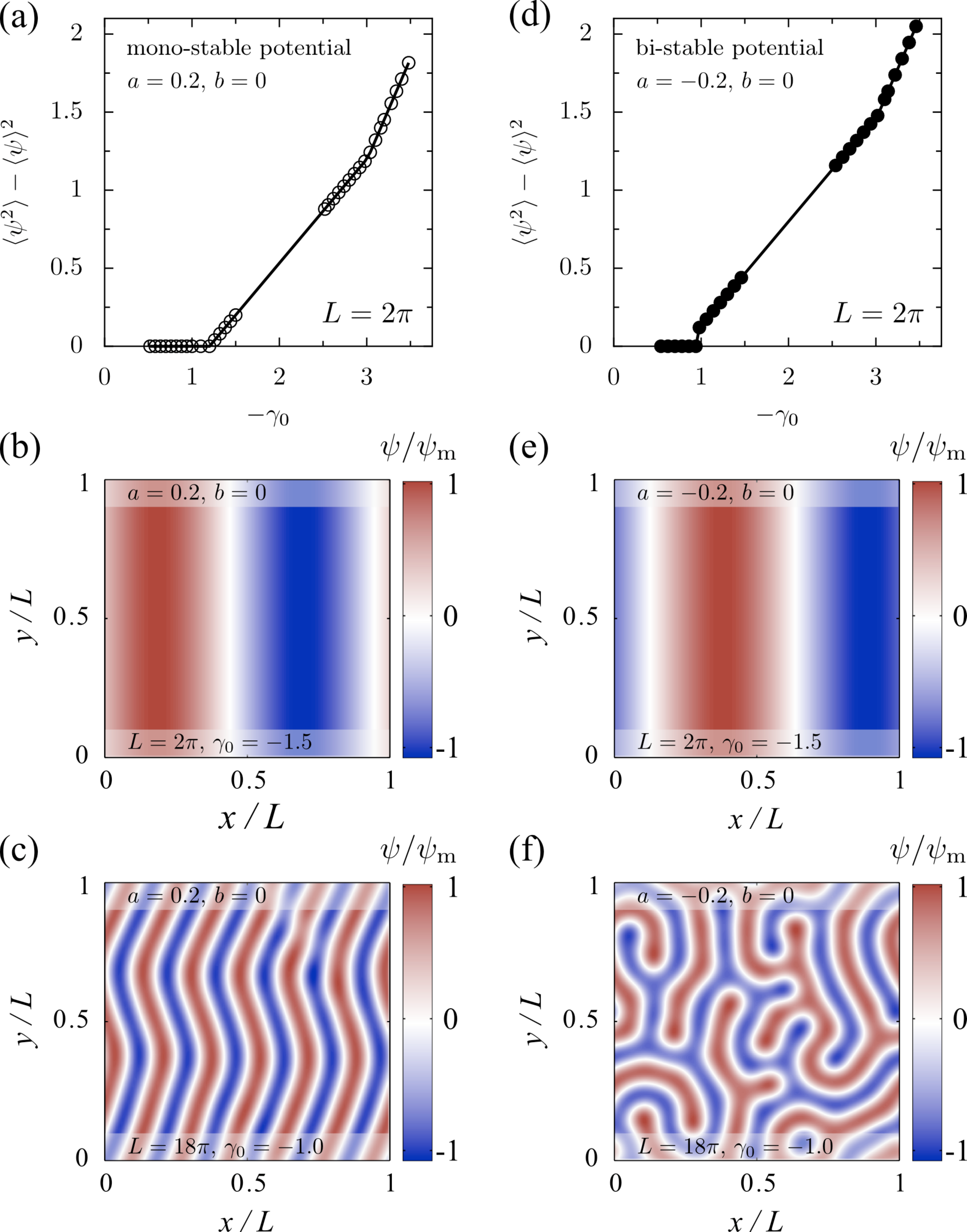} 
  \caption{Numerical illustration of  structural transitions in the order-parameter $\psi$ for (a-c) mono-stable and (d-f) bi-stable potentials. (a,d) Symbols show the results of simulations for the first two $\gc_0$-induced transitions, and lines are linear interpolations.  Quasi-stationary space-time averages $\langle\,\cdot \,\rangle$ were computed over $3000$ successive simulation time-steps ($\Delta t=0.1$) after an initial relaxation period of $200$ characteristic time units $t_\mathrm{u}=L^4/\gc_2$. (b,c)~Snapshots of the order-parameter field $\psi$ at $t=500$,  scaled by the maximum value $\psi_\mathrm{m}$, for a mono-stable potential $U(\psi)$ and homogeneous random initial conditions.  After the first transition two stripes appear, and the number of stripes increases with the number of transitions. (e,f) Snapshots of the order-parameter at $t=500$ for a bi-stable potential.  For $\gc_0\ll -(2\pi)^2\gc_2/L^2$, increasingly more complex quasi-stationary structures arise;  see References~\cite{2006Aranson,1997Swinney} for similar patterns in excited granular media and chemical reaction systems.
 \label{fig01_psi}
 }
\end{figure}

\paragraph{Structural transitions}
Results from the numerical simulations for the order-parameter field~$\psi(t,\mathbf x)$ and two qualitatively different potentials $U(\psi)$ are summarized in~\Fref{fig01_psi}. In these simulation, the parameter $\gamma_0$ was varied between successive runs while keeping all other parameters fixed.
To quantify  changes in the quasi-stationary dynamics of~$\psi$ as a function of $\gamma_0$, we measured the space-time averaged standard deviation \mbox{$\gs_\psi^2=\langle {\bf \psi} ^2 \rangle - \langle {\bf \psi} \rangle^2$}  (\Fref{fig01_psi}a,d).  Regions with $\gs_\psi^2=0$ correspond to  disordered structureless stationary states, whereas $\gs^2_\psi>0$ indicates the emergence of stationary or quasi-stationary dynamical structures. Singular points in the curve $\gs^2(\gc_0)$ signal qualitative changes in the order-parameter dynamics.
\par 
In the case of a mono-stable potential ($a>0$), the quantifier $\gs^2_\psi$ undergoes a series of  continuous transitions as $\gc_0$ is lowered to negative values, see \Fref{fig01_psi}a. Each of those transitions corresponds to an increase in the number of \lq stripes\rq~that are found to persist for long periods of time in the simulations (\Fref{fig01_psi}b,c). By contrast, in the case of bi-stable potentials ($a<0$), the onset of pattern formation carries the signature of a first-order transition reflected by a sudden jump in $\gs^2_\psi$ (\Fref{fig01_psi}d). However, while such singularities in~$\gs^2_\psi$  share some formal similarities with macroscopic phase transitions, one could also argue that they merely signal a change in the typical number of excitable modes in the system. In fact, by viewing such elementary excitations as \lq quasi-particles\rq, the structural transitions in \Fref{fig01_psi}a,d appear to be more closely related to finite-systems singular points~\cite{2006DuHi,2006HiDu,2008Kastner}.
\par
An estimate of the critical absolute value $\gc_0$ for the first \lq disorder-structure\rq~transition can be obtained by dimensional analysis, or  by equating the last two terms in Equations~\eref{e:psi_0} or~\eref{e:psi_pm}, yielding  $\gc^c_0\approx- (2\pi)^2\gc_2/L^2$. For $\gc_0\ll \gc_0^c$, increasingly more  complex  quasi-stationary patterns may arise (\Fref{fig01_psi}c,d). Structures similar to those in \Fref{fig01_psi} have been observed and widely studied~\cite{2009Cross} in granular media~\cite{2006Aranson} and chemical systems~\cite{1997Swinney}. In the next section,  we shall demonstrate that certain aspects of collective microbial motion can be described within the same  class of fourth-order PDEs.

\begin{figure}[b]
\centering
  \includegraphics[width=12.5cm]{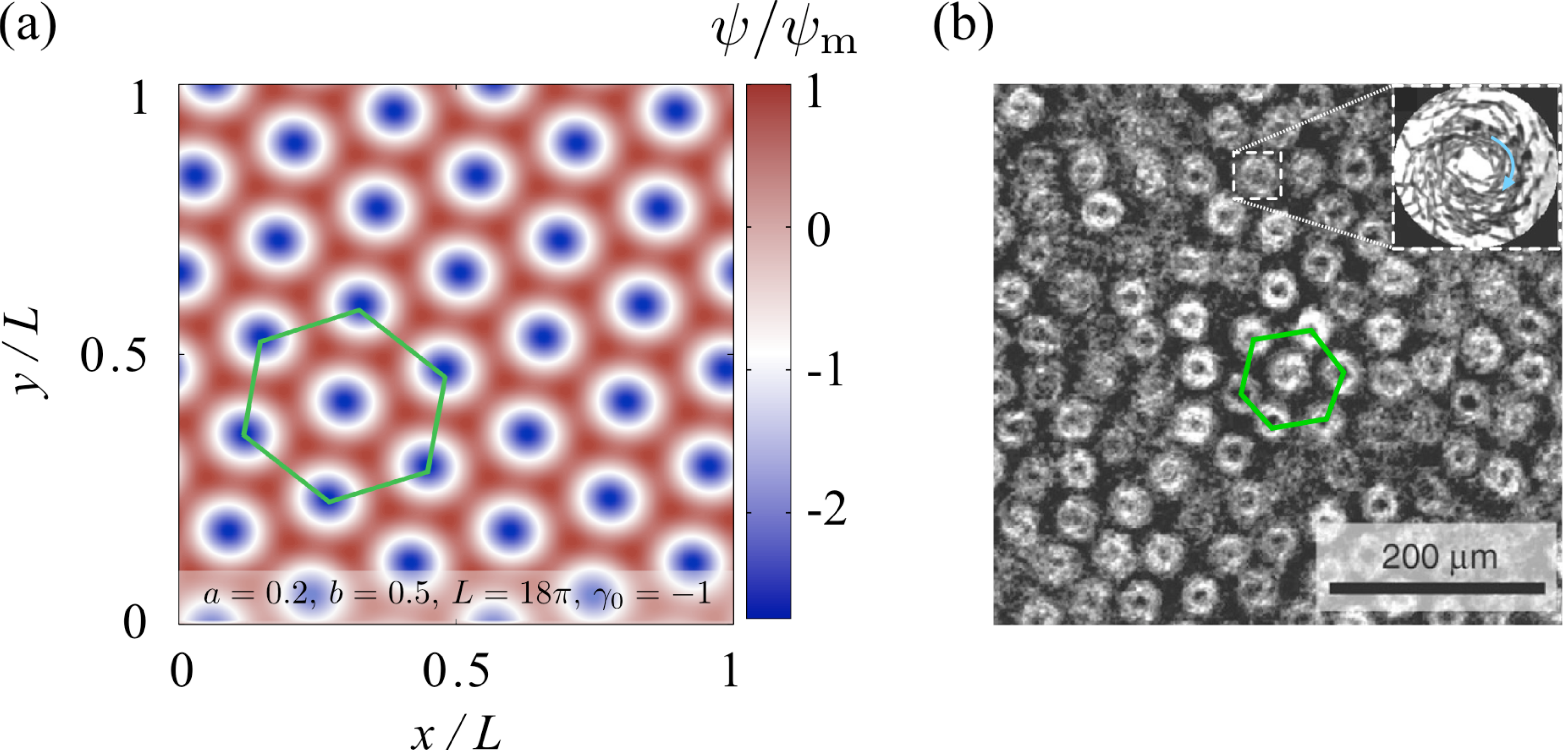} 
  \caption{Effect of symmetry breaking. (a) Stationary hexagonal lattice of the pseudo-scalar vorticity order-parameter $\psi=\omega$,  scaled by the maximum value $\psi_\mathrm{m}=\omega_\mathrm{m}$,   as obtained in simulations of Equations~\eref{e:scalar} and~\eref{e:scalar_potential} with \mbox{$b >0$}, corresponding to a broken reflection symmetry $\omega\not\to-\omega$. Blue regions correspond to clockwise motions. (b)  Hexagonal vortex lattice formed spermatozoa of sea urchins (\textit{Strongylocentrotus droebachiensis}) near a glass surface; from~\cite{2005Riedel_Science}  adapted and reprinted with permission from AAAS. At high densities, the spermatozoa assemble into vortices that rotate in clockwise direction (inset)  when viewed from the bulk fluid.
  \label{fig02_ingmar}}
\end{figure}

\subsection{Symmetry breaking}
\label{s:symmetry_breaking}

With regard to microbial suspensions, the minimal model~\eref{e:scalar} is useful for illustrating how microscopic symmetry-breaking mechanisms that affect the motion of individual organisms or cells~\cite{2005Berg_Nature,2008Tang_PNAS,2010Elgeti_BiophysJ,2012Dunstan}  can be implemented into macroscopic field equations. To demonstrate this,  we interpret $\psi$ as a 2D pseudo-scalar vorticity field\footnote{$\eps_{ij}$ denotes the Cartesian components of the Levi-Civita tensor, $\partial_i=\partial/\partial x_i$ for $i=1,2$, and we use a summation convention for equal indices throughout.} 
\be
\psi\equiv \omega=\nabla\wedge \mathbf v=\eps_{ij}\partial_iv_j,
\ee
which is assumed to describe the  flow dynamics $\mathbf v$ of a dense microbial suspension  confined to a thin quasi-2D layer of fluid. If the confinement mechanism is top-bottom symmetric, as for example  in a  thin free-standing bacterial film~\cite{2007SoEtAl}, then one would expect that vortices of either handedness are equally likely. In this case, \Eref{e:scalar} must be invariant under $\go\to -\go$, implying that $U(\go)=U(-\go)$ and, therefore, $b=0$ in \Eref{e:scalar_potential}. Intuitively, the transformation  $\go\to -\go$ corresponds to a reflection of the observer position at the midplane of the film (watching the 2D layer from above~\textit{vs.} watching it from below). 
\par
The situation can be rather different, however, if we consider the dynamics of microorganisms close to a liquid-solid interface, such as the motion of bacteria or sperms cells in the vicinity of a glass slide (\Fref{fig02_ingmar}). In this case, it is known that the trajectory of a swimming cell can exhibit a preferred handedness~\cite{2005Berg_Nature,2008Tang_PNAS,2010Elgeti_BiophysJ,2012Dunstan}. For example, the bacteria \textit{Escherichia coli}~\cite{2005Berg_Nature}  and \textit{Caulobacter}~\cite{2008Tang_PNAS} have been observed to swim in circles when confined near to a solid  surface. More precisely,  due to an intrinsic chirality in their swimming apparatus,  these organisms move on circular orbits in clockwise (anticlockwise) direction when viewed from inside the bulk fluid (glass surface).  {\color{black}{Qualitatively  similar behavior has also been reported for sea urchin sperm swimming close to solid surfaces~\cite{1980Gibbins_Circle}.  
}} 
\par
Hence, for various types of swimming microorganisms, the presence of the near-by no-slip boundary  breaks the reflection symmetry,  $\go\not\to -\go$. The simplest way of accounting for this in a macroscopic continuum model is to adapt the potential $U(\go)$ by permitting values $b\ne 0$ in~\Eref{e:scalar_potential}. The result of a simulation with $b>0$ is shown in~\Fref{fig02_ingmar}a. In contrast to the symmetric case $b=0$ (compare \Fref{fig01_psi}c), an asymmetric potential favors the formation of stable hexagonal vorticity patterns ~(\Fref{fig02_ingmar}a) --  such self-assembled hexagonal vortex lattices have indeed been observed experimentally~\cite{2005Riedel_Science}  for  highly concentrated spermatozoa of sea urchins (\textit{Strongylocentrotus droebachiensis}) near a glass surface~(\Fref{fig02_ingmar}b).

\section{Vector model for an incompressible active fluid}
\label{sec:vector}

We now generalize the preceding considerations to identify a minimal vector-field model   for dense microbial suspensions. Previously developed continuum theories~\cite{2010Ramaswamy,2005ToTuRa,1998TonerTu_PRE,2008Wolgemuth,2009BaMa_PNAS,2010Pedley,2002Ra,2008SaintillanShelley} of microbial fluids typically distinguish solvent concentration, bacterial density, solvent velocity, bacterial velocity, and various orientational order-parameter fields (polarization, $\mathbf Q$-tensors, etc.). Aiming  to identify a minimal  hydrodynamic model,  we  construct  a simplified higher-order theory by focussing exclusively on the dynamics of the mean bacterial\footnote{Note that, while the joint momentum of a bacteria-solvent mixture~\cite{2012Marchetti} is conserved, the dynamics of the active (bacterial) component alone, as considered here, does not satisfy such a  conservation law.} velocity field $\mathbf v(t,\mathbf x)$ and restricting ourselves to the incompressible limit. By construction, the resulting $\mathbf v$-only theory, which is essentially a minimal Swift-Hohenberg-type~\cite{1977SwiftHohenberg} extension of the Toner-Tu model~\cite{2005ToTuRa,1998TonerTu_PRE}, may not be applicable to swarming or flocking regimes, where density fluctuations are dominant, but it can provide a useful basis for quantitative comparisons with experiments and simulations on highly concentrated active suspensions~\cite{2012Wensink}.  In practice, $\mathbf v$ can be determined applying suitable coarse-graining procedures (PIV algorithms, local averaging, etc.) to discrete experimental or numerical velocity data~\cite{2012Wensink,2003Bazant}.

\subsection{Model equations} 
Postulating incompressibility, {\color{black}{which is a good approximation for very dense suspensions~\cite{2012Wensink},}} 
\be\label{e:incompressibility}
\label{e:div_free}
\nabla \cdot \mathbf v=\partial_iv_i=0,
\ee
we assume that the dynamics of $\mathbf v$ is governed by the generalized Navier-Stokes equation
\begin{eqnarray}
(\partial_t + \mathbf v\cdot \nabla) \mathbf v=
-\nabla p  - (A+C |\mathbf v|^2)\mathbf v + \nabla\cdot \mathbf E.
\label{e:conti-b}
\end{eqnarray}
The pressure $p(t,\mathbf x)$ is the Lagrange multiplier for the incompressibility  constraint. Similar to the scalar case, \Eref{e:scalar_potential} above,  the $(A, C)$-terms in Equation~\eref{e:conti-b}  represent a quartic Landau velocity potential~\cite{2010Ramaswamy,2005ToTuRa,1998TonerTu_PRE}
\be\label{e:U_v}
U(\mathbf v)= \f{A}{2} |\mathbf v|^2+ \f{C}{4}|\mathbf v|^4. 
\ee
 {\color{black}{
 Physically, the inclusion of a polar ordering potential accounts for the fact that microorganisms typically exhibit head-tail asymmetries that may favor polar alignment, as manifested in the \lq bionematic\rq\space jets that form in bacterial suspensions~\cite{2007Cisneros,2011Cisneros_PRE}.  For $A>0$ and $C>0$, the potential is mono-stable and the fluid is damped towards a disordered state with $\mathbf v=0$. By contrast, for $A<0$, \Eref{e:U_v}  describes a $d$-dimensional  mexican-hat (sombrero) potential with fixed-points $|\mathbf v|=\sqrt{-A/C}$ corresponding to global polar order.  However, the fact that polar ordering appears only  locally but not globally in suspensions of swimming bacteria~\cite{2004DoEtAl,2007Cisneros,2011Cisneros_PRE} suggests that other instability mechanisms must be at work~\cite{2002Ra}. To capture this mathematically, one must either introduce additional order parameters~\cite{2010Ramaswamy,2005ToTuRa,1998TonerTu_PRE} or  destabilize the theory by identifying a suitable phenomenological ansatz for the effective stresses~\cite{1977SwiftHohenberg}.  
 }}
 Adopting the latter approach, we postulate that the components of the symmetric and traceless rate-of-strain $\mathbf E$ tensor are given by
\be\label{e:E}
E_{ij} =
\Gc_0(\partial_i v_j +\partial_j v_i)  -
\Gc_2\triangle\,(\partial_i v_j +\partial_j v_i)  +
S\, q_{ij},
\ee
where 
\be\label{e:q_ij}
q_{ij}= {v_i v_j}-\f{\gd_{ij}}{d}|\mathbf v|^2
\ee
is a $d\times d$-dimensional mean-field approximation to the $\mathbf Q$-tensor, representing active {\color{black}{nematic stresses~\cite{2002Ra,2008BaMa}} }due to swimming ($\gd_{ij}$ is the Kronecker tensor).  Although the $S$-term does not affect the linear stability of the model, general hydrodynamic arguments~\cite{2010Pedley} imply that $S<0$ for pusher-swimmers  like \emph{E. coli}~\cite{2011DrescherEtAl} or \emph{B. subtilis}, whereas $S>0$ for puller-type microswimmers such as  \textit{Chlamydomonas} algae~\cite{2010DrEtAl_PRL}.  {\color{black}{The $\Gamma_0$-term in~\eref{e:E} is dictated by the requirement that the model contains the Navier-Stokes equations as a limit case, and the $\Gc_2$-damping term is motivated by generic stability considerations, as recent experiments~\cite{2012Wensink} suggest that $\Gc_0$ can become negative in dense bacterial suspensions.}}
Inserting Equations~\eref{e:E} and~\eref{e:q_ij} into \Eref{e:conti-b}, and defining
\be
\gl_0=1-S, \qquad \gl_1=-S/d,
\ee
 we obtain 
 \be
 \fl
\qquad\quad
 (\partial_t + \gl_0 \mathbf v\cdot \nabla) \mathbf v
 =
-\nabla p  +\gl_1 \nabla \mathbf v^2 - (A+C |\mathbf v|^2)\mathbf v + 
\Gc_0 \triangle \mathbf v -\Gc_2\triangle^2    \mathbf v.
\label{e:conti-c}
 \ee
 {\color{black}{A variational formulation of the combined field equations~\eref{e:div_free} and~\eref{e:conti-c} is given in \ref{app:vector}.}}
  \par
 For  $\Gc_0>0$ and $\Gc_2=0$, \Eref{e:conti-c} reduces to an incompressible version of the classical Toner-Tu model ~\cite{2010Ramaswamy,2005ToTuRa,1998TonerTu_PRE}. It is, however,  the combination  of the two $\Gc$-terms  with the non-variational convective derivative that turns out to be crucial for the formation of self-sustained quasi-chaotic flow patterns. The linear $\Gc$-terms are reminiscent of  the higher-order spatial derivatives in the classical Swift-Hohenberg theory~\cite{1977SwiftHohenberg},  see~\Eref{e:scalar}, and  \Eref{e:conti-c} with $\Gc_0<0$ and $\Gc_2>0$ yields a simple -- if not the simplest -- generic continuum description of turbulent meso-scale instabilities observed in dense bacterial suspensions~\cite{2012Wensink}. More generally, \Eref{e:conti-c} can provide a satisfactory phenomenological model whenever interaction terms in more complex field theories, that lead to instabilities in the $\mathbf v$-field,  can be effectively approximated by a fourth-order Taylor expansion in Fourier space. This is likely to be the case for a wide range of active systems. Phrased differently,  the last two terms in~\Eref{e:conti-c} may be regarded as the Fourier-space analogue of the Toner-Tu driving terms, which correspond to a series expansion in terms of the order-parameter.  Hence, similar to the higher-order gradient terms in the  scalar theory from \Eref{e:scalar},  the $(\Gc_0,\Gc_2)$-terms in \Eref{e:conti-c} describe intermediate-range interactions, and their role  in Fourier-space is similar to that of the Landau potential in velocity space.

\subsection{Linear Stability Analysis}
To support the qualitative statements in the preceding paragraph, we now perform a stability analysis for the 2D case relevant to the simulations discussed below, assuming $\Gc_0<0$ and $C>0$, $\Gc_2>0$.  
\par
The fixed points of Equations~\eref{e:div_free} and \eref{e:conti-c} are given by the extrema of the quartic velocity potential $U(\mathbf v)$. For arbitrary values of~$A$, Equations~\eref{e:div_free} and \eref{e:conti-c} have a fixed point that corresponds to a disordered isotropic state $(\mathbf v,p)=(\mathbf 0,p_0)$ where  $p_0$ is a constant pressure. For $A<0$, an additional class of fixed points arises, corresponding to a manifold of globally ordered polar states $(\mathbf v,p)=(\mathbf v_0,p_0)$, where $\mathbf v_0$ is constant vector with arbitrary orientation and fixed swimming speed $|\mathbf v_0|=\sqrt{-A/C}=:v_0$.
\par
Linearizing Equations~\eref{e:div_free} and \eref{e:conti-c}  for small velocity and pressure perturbations around the isotropic state, $\mathbf v=\mathbf{\epsilon}$ and $p=p_0+\eta$ with $|\eta|\ll |p_0|$,  and considering perturbations of the form
\be
(\eta,\mathbf \epsilon)=(\hat\eta,\hat{\mathbf \epsilon}) \exp(- \gs_0 t-i\mathbf k\cdot \mathbf x),
\ee
we find
\begin{eqnarray}
0  &=& \mathbf k\cdot\hat{\mathbf \epsilon},
\\
\gs_0 \hat{\mathbf \epsilon} & =& - i\hat \eta \mathbf k 
+ (A+ \Gamma_0 |\mathbf k|^2 + \Gamma_2 |\mathbf k|^4)\hat{\mathbf \epsilon}.
\qquad
\end{eqnarray} 
Multiplying the second equation by $\mathbf k$ and using the incompressibility condition implies that $\hat{\eta}=0$ and, therefore,
\be\label{e:spec_iso}
\gs_0(\mathbf k)= A+ \Gamma_0 |\mathbf k|^2 + \Gamma_2 |\mathbf k|^4.
\ee
Assuming $\Gc_0<0$ and $\Gc_2>0$,  and
provided that $
4A<|\Gc_0|^2/\Gc_2,
$
we find an unstable band of modes with $\gs_0(\mathbf k)<0$ for $
k_-^2<|\mathbf k|^2< k_+^2$, where
\be
k_\pm^2=
\f{|\Gc_0|}{\Gc_2}
\left(\f{1}{2}\pm \sqrt{\f{1}{4}-\f{A\Gamma_2 }{ |\Gamma_0|^2 }}
\right).
\ee
For $A<0$ the isotropic state is generally unstable with respect to long-wavelength (i.e., small-$|\mathbf k|$)  perturbations.

\par
We next perform a similar analysis for the polar state~$(\mathbf v_0,p_0)$, which is energetically preferred for~$A<0$ and corresponds to all active particles swimming in the same direction (\lq global order\rq). In this case, when considering small deviations
\be
\mathbf v=\mathbf v_0+ \mathbf \epsilon,
\qquad
p=p_0+\eta,
\ee
it is useful to distinguish perturbations perpendicular and parallel to $\mathbf v_0$, by writing $\mathbf \epsilon= \mathbf \epsilon_{||} +  \mathbf \epsilon_\perp$ where $\mathbf v_0\cdot  \mathbf \epsilon_\perp=0$ and $\mathbf v_0\cdot  \mathbf \epsilon_{||}=v_0\epsilon_{||}$. Without loss of generality, we may choose $\mathbf v_0$ to point along the $x$-axis, $\mathbf v_0=v_0\mathbf e_x$. Adopting this convention, we have $\mathbf \epsilon_{||}=(\epsilon_{||}, 0)$ and  $\mathbf \epsilon_\perp=(0,\epsilon_\perp)$, and to leading order
\be
|\mathbf v|^2\simeq v_0^2+2 v_0 \epsilon_{||}.
\ee
Linearization for exponential perturbations of the form
\be
(\eta,\epsilon_{||},\epsilon_{\perp} )=
(\hat{\eta},\hat{\epsilon}_{||},\hat{\epsilon}_{\perp}) \,\exp(-\gs t -i\mathbf k\cdot \mathbf x)
\ee
yields 
\be
0  &=& \mathbf k\cdot\hat{\mathbf \epsilon},
\label{e:perturbation-2.0-incomp}\\
\gs \; \hat{\mathbf \epsilon }
&=&
-i (\hat{\eta}  - 2v_0\gl_1 \hat \epsilon_{||}) \mathbf k
-
\mathbf M\hat{\mathbf\epsilon},
\label{e:perturbation-2.0}
\ee
where
\be
\mathbf{M}=
\left(\begin{array}{cc}
2A  & 0\\
0  &  0
\end{array}\right)
-(\Gamma_0 |\mathbf k|^2+ \Gamma_2 |\mathbf k|^4-i\gl_0 k_x v_0)\mathbf I 
\ee
with $\mathbf I=(\gd_{ij})$ denoting the identity matrix.
Multiplying Equation~\eref{e:perturbation-2.0} with $i\mathbf k$, and using the incompressibility condition~\eref{e:perturbation-2.0-incomp},  gives 
\be
\hat{\eta}= 
2v_0\gl_1  \epsilon_{||} + i\f{\mathbf k\cdot(  \mathbf M\hat{\mathbf\epsilon})}{|\mathbf k|^2}.
\label{e:perturbation-polar-eta}
\ee
Inserting this into Equation~\eref{e:perturbation-2.0} and defining $\mathbf M_\perp=\mathbf \Pi(\mathbf k) \mathbf M$, where
\be
\Pi_{ij}(\mathbf k)=\gd_{ij} -\f{k_ik_j}{|\mathbf k|^2}
\ee
is the orthogonal projector of $\mathbf k$, we obtain 
\be
\gs \, \hat{\mathbf \epsilon} 
=
- \mathbf M_\perp\, \hat{\mathbf\epsilon}.
\ee
The eigenvalue spectrum of the matrix $\mathbf M_\perp$ is given by
\be\label{e:spec_polar}
\gs(\mathbf k)\in \left\{0, \left(\Gamma_0 |\mathbf k|^2+ \Gamma_2 |\mathbf k|^4-2A\f{k_x^2}{|\mathbf k|^2}\right)-i \gl_0 v_0  k_x\right\}.
\ee
The zero eigenvalues correspond to the Goldstone modes. The non-zero eigenvalues have eigenvectors $(-k_y,k_x)$, implying that, for $\Gc_0<0$, there will be a range of exponentially growing  modes in the direction perpendicular to $\mathbf k$. 
\par
Equations~\eref{e:spec_iso} and~\eref{e:spec_polar} predict that, when $A<0$ and $\Gc_0<0$,  isotropic and polar fixed points become simultaneously unstable, thereby signaling the existence of spatially inhomogeneous dynamic attractors. More generally,  within the class of standard PDEs,  the  two $\Gc$-terms in \Eref{e:conti-c} appear to provide the simplest \lq linear way\rq~of obtaining a $\mathbf v$-only theory that exhibits non-trivial stationary dynamics.
In principle, one could also try to model instabilities by combining odd or fractional powers of $|\mathbf k|$ in Equations~\eref{e:spec_iso} and~\eref{e:spec_polar}; this would be analogous to replacing the quartic Landau potential by a more general function of $|\mathbf v|$. However, when considering eigenvalue spectra based on odd or non-integer powers of $|\mathbf k|$,  the underlying dynamical equations in position space would become fractional PDEs. Such fractional models could potentially be useful for describing active suspensions with long-range or other types of more complex interactions, but their analysis goes far beyond the scope of this paper.

\subsection{Numerical results in 2D}

We simulated the vector model,  defined by Equations~\eref{e:incompressibility} and \eref{e:conti-c}, in two space-dimensions using an algorithm similar to that described  in Section~\ref{s:scalar_numerics}. The primary difference compared with the simulations for the scalar model is an additional pressure correction subroutine that ensures the incompressibility of the flow (see  Reference~\cite{2012Wensink} for details). Table~\ref{para_vec} summarizes characteristic units and rescaled parameters as adopted in the  computations.
All  simulations were initiated with random initial conditions, and the typical time discretization was $\Delta t=0.1$ (in characteristic time units  $t_\mathrm{u}$).

\begin{table}[b]
\centering
\begin{tabular}{c|c}
model parameter & rescaled dimensionless parameter \\
\hline
$A$ &  $ A \;t_\mathrm{u}$\\
$C$ & $1$ \\
\hline
$\Gamma_0$ & $  \Gamma_0\;t_\mathrm{u}L^{-2}$ \\
$\Gamma_2$ & $1$\\
$ S$ & $  S $
\end{tabular}
\caption{The right column shows the rescaled dimensionless parameters used in the simulations of the vector model from \Eref{e:conti-c}. The unit time is defined by the damping time-scale $t_\mathrm{u} =L^4 /\Gamma_2$, where $L$ is the length of the 2D simulation box,  and the order-parameter is measured in units of  $v_\mathrm{u} =1/\sqrt{t_\mathrm{u}C}$.
 \label{para_vec}}
\end{table}

\begin{figure}[t]
\centering
 \includegraphics[width=13.5cm]{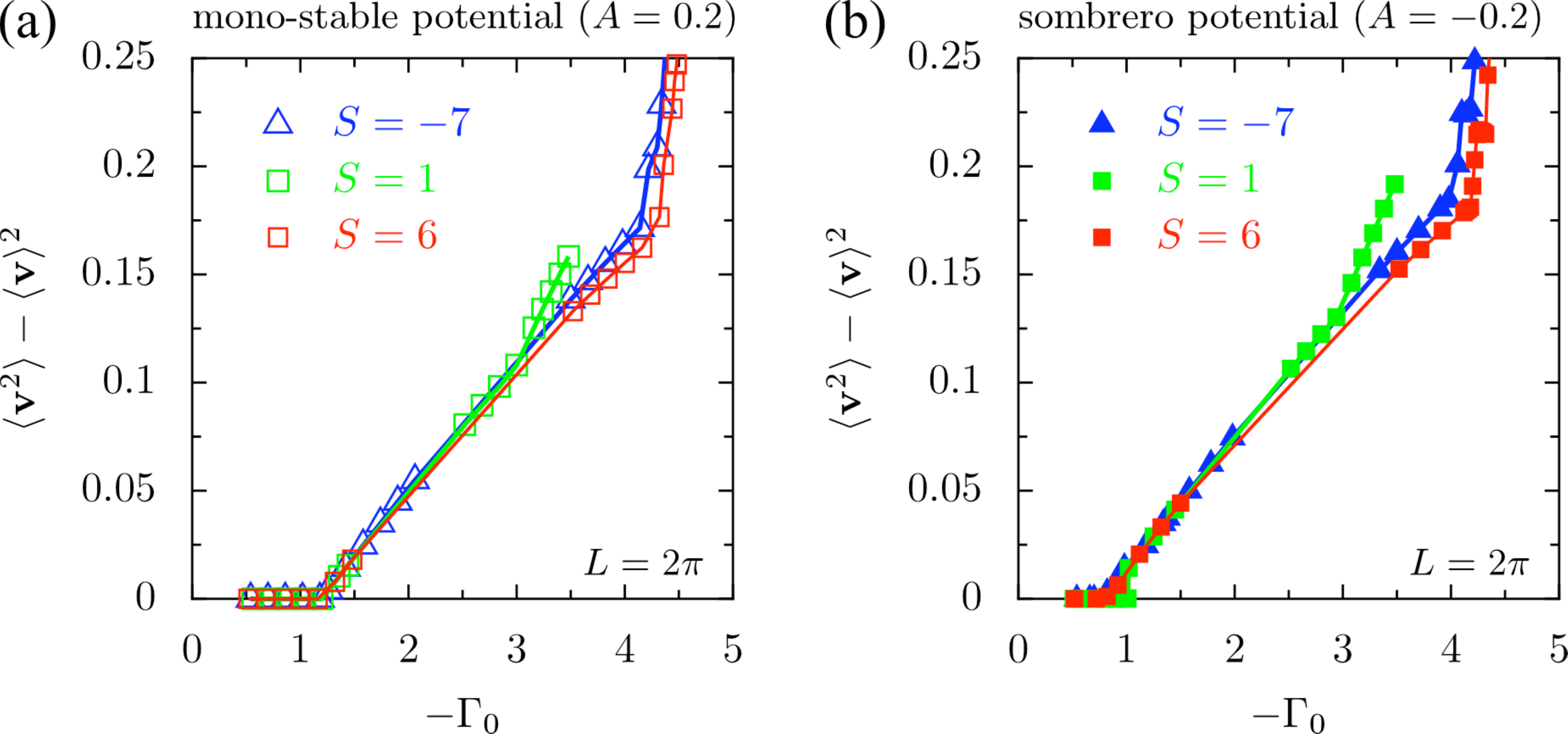} 
\caption{First few kinematic transitions in the quasi-stationary dynamics of the vector model for (a) mono-stable and (b) mexican-hat potentials at different values of $S$.  Transitions are indicated by sudden changes of order-parameter variance $\gs_v^2=\langle {\bf v} ^2 \rangle - \langle {\bf v} \rangle^2$. The space-time averages $\langle\,\cdot\,\rangle$ were measured from $3000$ successive snapshots $(\Gd t=0.1)$ after an initial relaxation period of $200$ time units $t_\mathrm{u}$. 
\label{fig03_v}}
\end{figure}

\paragraph{Kinematic transitions}
We first study how a decrease of the \lq viscosity\rq\space parameter~$\Gamma_0$ affects the stationary dynamics for  mono-stable and mexican-hat (polar-ordering) potentials. To this end, simulations were  performed with fixed potential functions at three different values of the pusher/puller parameter $S$, while varying $\Gamma_0$ between successive runs. Changes in the quasi-stationary dynamics are quantified by measuring the space-time averaged variance \mbox{$\gs_v^2=\langle \mathbf v^2\rangle-\langle \mathbf v\rangle^2$}, shown in  \Fref{fig03_v}.  Similar to the scalar model, the first transition  from an isotropic state with $\mathbf v\equiv0$ to a non-trivial stationary dynamics with $\gs_v^2>0$ is found to occur at $\Gc_0\approx -(2\pi)^2\Gamma_2/L^2$. 

\begin{figure}[t]
\centering
 \includegraphics[width=12.7cm]{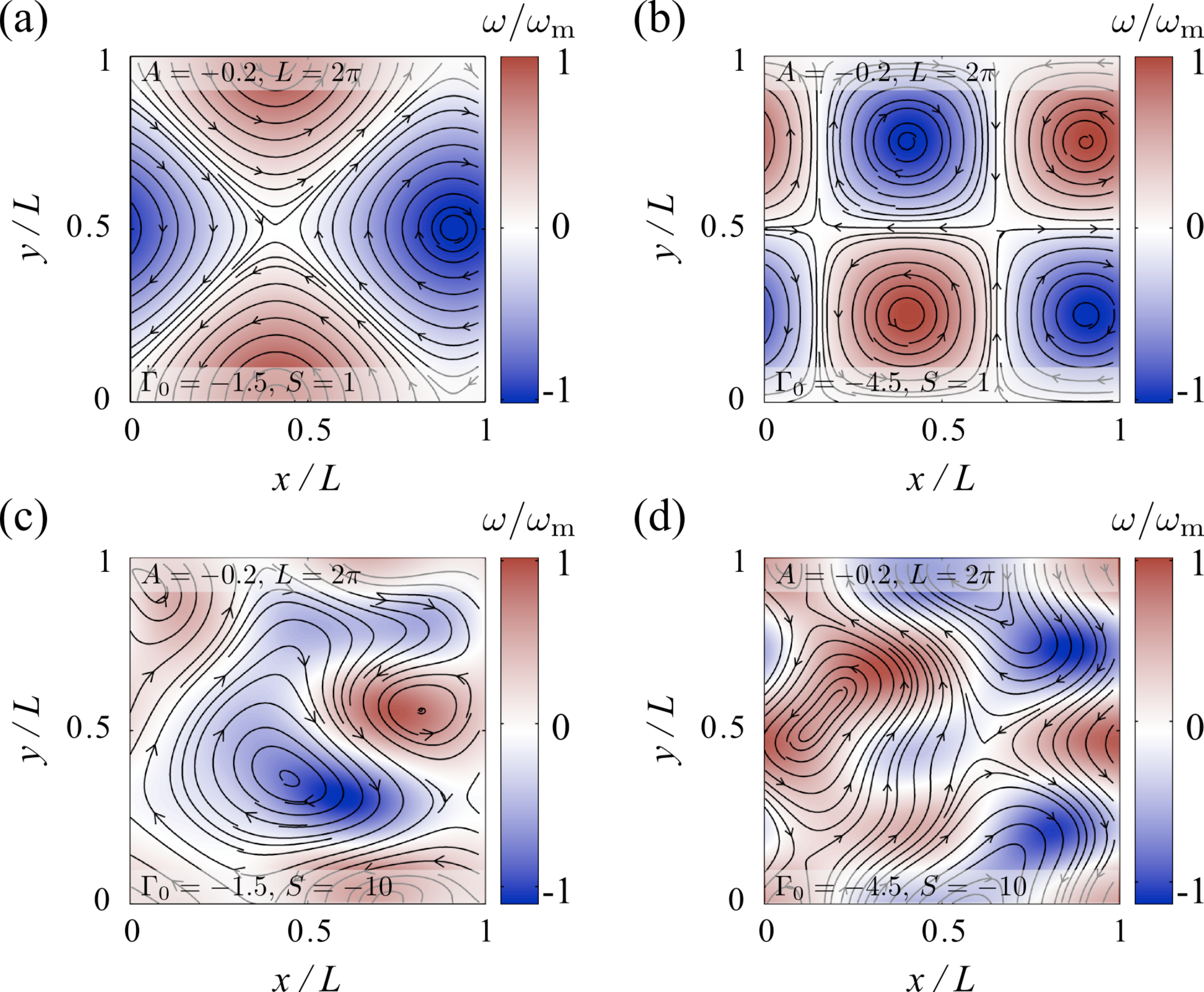} 
\caption{Simulation snapshots ($t=400$) of streamlines and vorticity (scaled by the maximum value $\omega_\mathrm{m}$)  for the sombrero (polar-ordering) potential and random initial conditions. (a,b)  Stationary vortex lattices  obtained in the special case $S=1$, corresponding to a vanishing convective derivative in~\Eref{e:conti-c}. (c,d)  Strong convection ($|S|\gg 1$) leads to the formation of dynamic patterns.
 \label{fig04}}
\end{figure}

To illustrate kinematic changes in the flow dynamics in more detail, we show quasi-stationary  snapshots from simulations with a mexican-hat potential for different values $(\Gc_0,S)$ in \Fref{fig04}. 
In the special case $S=1$,  corresponding to a vanishing convective derivative in~\Eref{e:conti-c}, stationary cubic vortex lattices form, with an increasing number of vortices as $\Gamma_0$ is decreased (\Fref{fig04}a,b). By contrast, for $S\ne 1$, nonlinear convective effects cause distortions in  the vortex lattices. As a consequence,  the dynamical system no longer approaches a time-independent stationary state but instead exhibits a complex non-equilibrium dynamics (\Fref{fig04}c,d). For pushers ($S<0$), the resulting turbulent flow patterns look remarkably similar  to those observed in dense suspensions of~\textit{Bacillus subtilis}~\cite{2004DoEtAl,2007Cisneros,2012Wensink,2011Cisneros_PRE}.

\begin{figure}[t]
\centering
 \includegraphics[width=12.7cm]{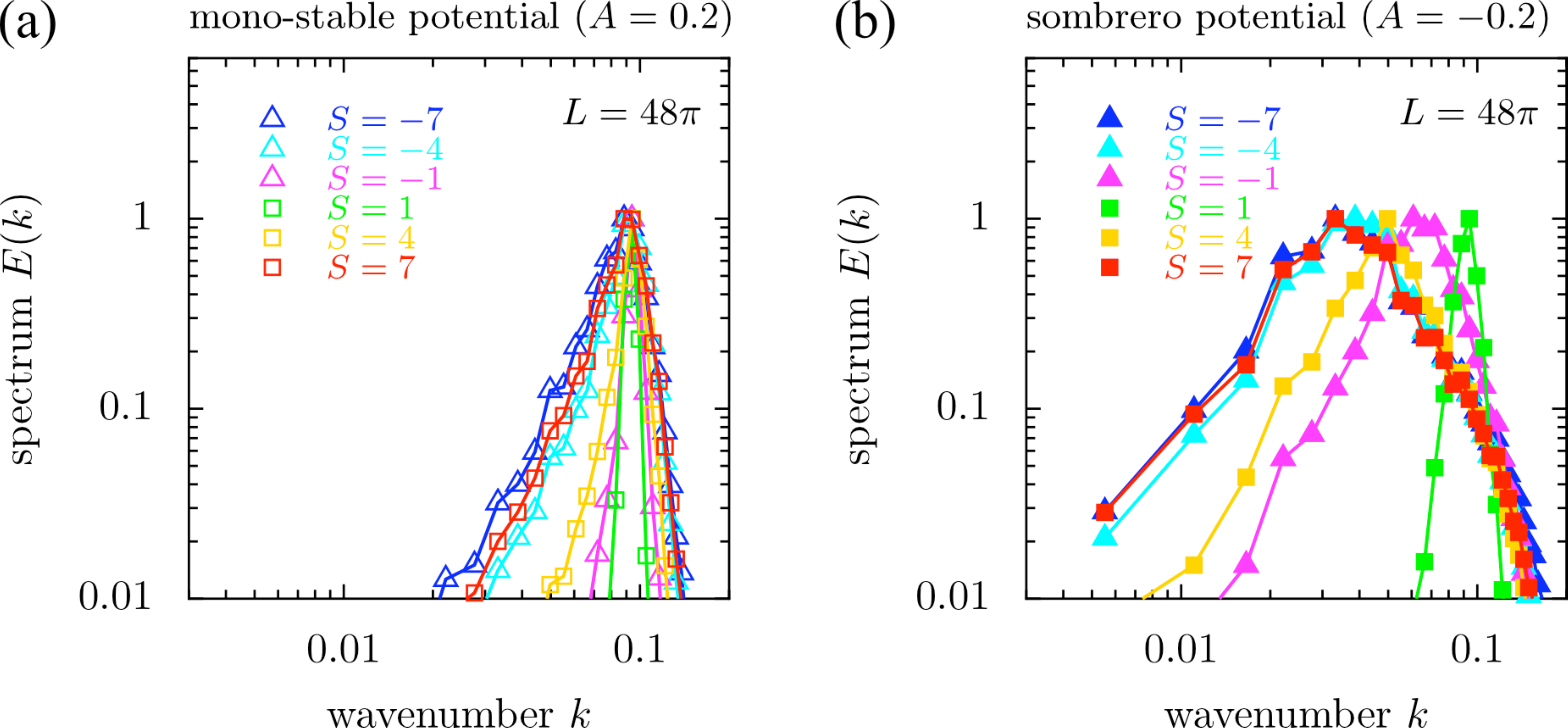} 
\caption{Flow energy spectra in arbitrary units for (a) mono-stable and (b) mexican-hat potentials and different values of $S$. Spectra are scaled to have the same maximal value. In the strongly non-linear regime, corresponding to $|S|\gg 1$,  the spectra of both pullers ($S>0$) and pushers ($S<0$) seem to approach \lq universal\rq\space limit functions, the exact shape of which depends on the type of the potential. Simulation parameters:  grid size $256\times 256$, $\Gamma_0 = -1$ in all simulations. 
\label{fig05_spectra}}
\end{figure}

\paragraph{Flow Spectra}
To resolve the structure of the flow fields obtained from Equations~\eref{e:incompressibility} and \eref{e:conti-c}, we calculated the energy spectrum $E(k)$, formally defined by 
\be
\langle \mathbf v^2 \rangle=2\int_0^\infty E(k) dk,
\ee 
where $k=|\mathbf k|$. By virtue of the Wiener-Khinchine theorem~\cite{2004Frisch}, $E(k)$ can be estimated by Fourier-transformation of the equal-time two-point velocity correlation function,  yielding in $d$ dimensions
\begin{equation} \label{e:spec}
E_d(k) \sim k^{d-1} \int d^d R\; e^{-i {\mathbf k} \cdot \mathbf R } \;
\langle {\mathbf v} (t,\mathbf r) \cdot {\mathbf v} (t,\mathbf r+\mathbf R)\rangle.
\end{equation}  
Traditionally,  spectral flow analysis has been an important tool in the investigation of classical turbulence phenomena~\cite{2004Frisch}. Flow spectra for our numerical data are summarized in~\Fref{fig05_spectra}. The critical case $S=1$ provides a useful reference point, since in this case the stationary dynamics becomes static, as illustrated in~\Fref{fig04}a,b for the sombrero potential. Accordingly, the spectra for $S=1$ exhibit a sharp peak that reflects the typical vortex size in the stationary state (see green curves in \Fref{fig05_spectra}a,b). In the case of the mono-stable potential,
changing the pusher/puller parameter to values $S\ne 1$ affects the asymptotic  slopes of the spectrum but leaves the position of the maximum practically constant (\Fref{fig05_spectra}a). By contrast,
for the polar-ordering potential, both slope and position of the maximum  change as $S$ is decreased or increased from unity (\Fref{fig05_spectra}b). The large-$|S|$ pusher-spectra for the mexican-hat potential agree well with those measured in dense quasi-2D \textit{B. subtilis}~\cite{2012Wensink} suspensions.

\begin{figure}
\centering
 \includegraphics[width=13.0cm]{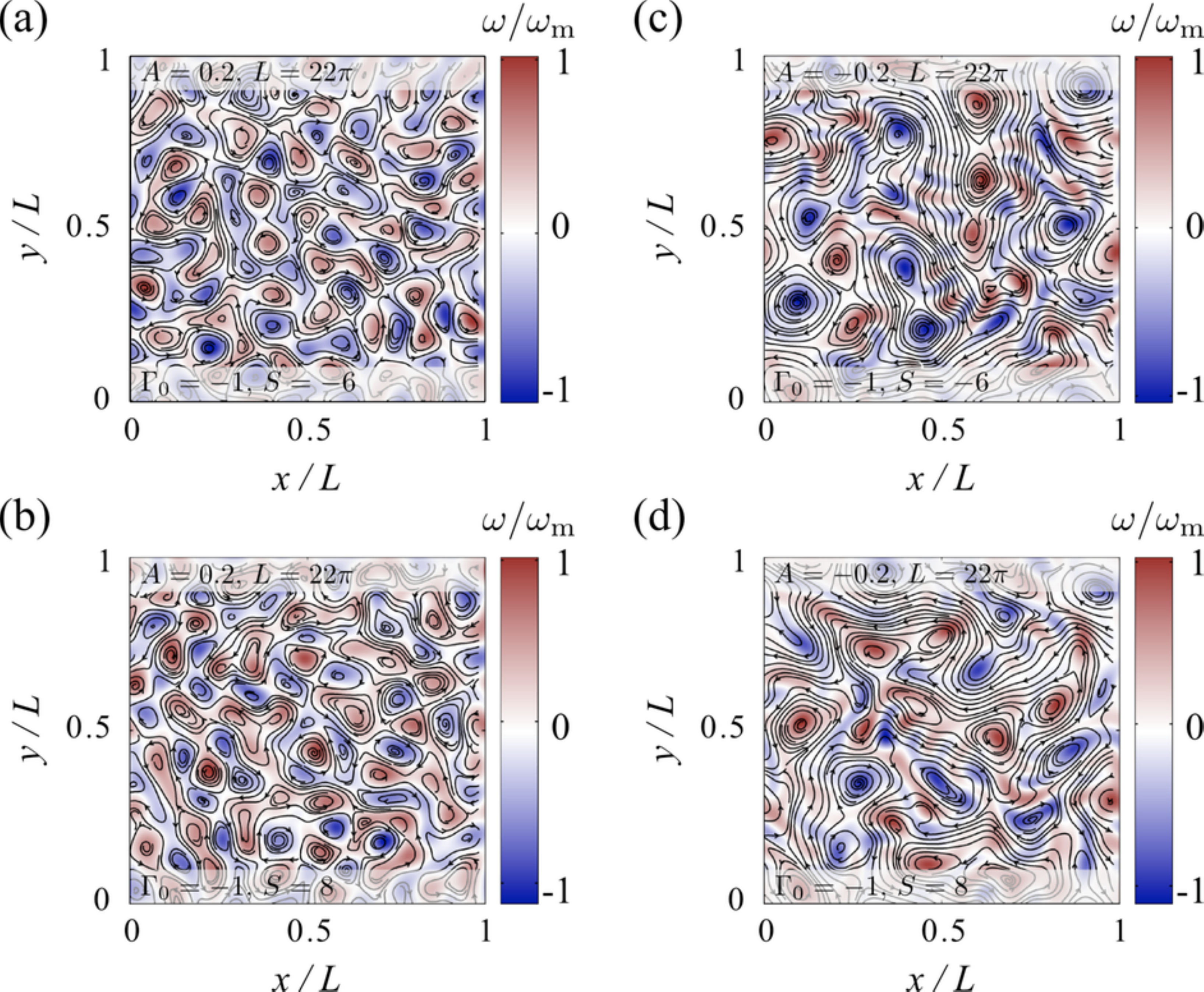} 
\caption{Snapshots of stationary flow and vorticity patterns, scaled by the maximum value $\omega_\mathrm{m}$,  for (a,b) mono-stable and (c,d) mexican-hat potentials. The pusher flow field in (c) agrees qualitatively with experimentally observed flow fields for dense \textit{B.~subtilis}~\cite{2004DoEtAl,2007Cisneros,2012Wensink,2011Cisneros_PRE} suspensions.
\label{fig06_v_snap_large}}
\end{figure}

\paragraph{Bionematic jets}
To illustrate qualitatively how the velocity potentials affect the stationary dynamics -- and, hence, the spectral flow properties --  we present in \Fref{fig06_v_snap_large} snapshots from computations with an intermediate-size  simulation volume and \mbox{$\Gc_0\ll -(2\pi)^2\Gc_2/L^2$}. For the mono-stable potential,  we observe fairly homogeneous vortex structures   (\Fref{fig06_v_snap_large}a,b) in agreement with the relatively sharp spectral peaks in  \Fref{fig05_spectra}a. By contrast,  in the case of the mexican-hat potential (\Fref{fig06_v_snap_large}c,d), extended jet-like regions form that look very similar  to the  \lq zooming bionematic\rq~phases in dense~\textit{B.~subtilis} suspensions~\cite{2004DoEtAl,2007Cisneros,2011Cisneros_PRE}.

\paragraph{Nonlinear limit \& \lq universality\rq}
Interestingly, our numerical results  suggest that  the spectra approach universal functional forms in the nonlinear limit $|\gl_0|=|S-1|\gg 1$, even though the exact asymptotic behavior depends on the type of the potential. Moreover,  for pushers with $\gl_0\gg 1$, the spectra obtained from~\Eref{e:conti-c} agree well  with those measured for dense quasi-2D  \textit{B.~subtilis} suspensions~\cite{2012Wensink}, and a 
very similar small-$k$ scaling was also observed in 2D particle simulations of self-propelled rods~\cite{2012Wensink}. These observations hint at some \lq universality\rq~in the spectral properties of dense active particle systems. Future investigations of this question will require larger simulations as well as systematic experimental investigations of larger systems, which allow to extract  asymptotic scaling laws and other spectral details with higher accuracy. Generally, in the strongly nonlinear regime $|\gl_0|=|1-S|\gg 1$, the vector model predicts phenomenologically similar behavior for both pullers ($S>0$) and pushers ($S<0$), see~\Fref{fig06_v_snap_large}. However, while values of $\gl_0\gg 1$ appear to have been realized in dense suspensions of  pusher-type \textit{B. subtilis} bacteria~\cite{2007Cisneros,2012Wensink,2011Cisneros_PRE}, it seems unclear at present  whether puller (e.g. algal) suspensions with $S\gg 1$ can be achieved in experiments.

\section{Conclusions and future challenges}

Identifying \lq universal\rq\space  features  of active fluids is one of the key challenges en route to a more systematic  physical classification of biological matter.  Here, we have argued that, in many cases,  phenomenological aspects of dynamical non-equilibrium phases can be naturally described by minimal models that focus on a limited number of experimentally accessible order-parameters and are based on higher-than-second-order spatial derivatives.  If true, this would imply that a variety of active fluids with different microscopic constituents can exhibit very similar long-wavelength behavior. More generally,  adopting the view that  kinematic transitions in living fluids reflect qualitative changes in small-wavenumber expansions, and thus may be interpreted as Landau-type transitions in Fourier space, can help to catalog non-equilibrium systems according to their asymptotic behavior at long  wavelengths, similar to the classification of phase transitions in  equilibrium thermodynamics.   
\par

Simplified continuum descriptions of active fluids remain practically relevant because many of the recently proposed multi-order-parameter theories feature a large number of unknown transport coefficients and, therefore, cannot be tested in detail with present and next generation data. While our above analysis has focussed on the most basic scalar and vector models, the approach can be easily extended to higher tensorial order-parameter fields (e.g., ${\mathbf Q}$-tensor descriptions~\cite{2010Mishra_JSM} of active nematics).  To test whether microscopically different active fluids do indeed share universal hydrodynamic long-wavelength characteristics will require combined analytical, numerical and  experimental efforts. First attempts to apply the above ideas to very dense  bacterial suspensions give promising results~\cite{2012Wensink} but further quantitative studies will be needed to decide whether  fourth and higher-order PDEs are capable of providing a sufficiently accurate phenomenological description of experimentally observed active phases. To expedite quantitative comparisons with experiments, it would be desirable to develop alternative computation schemes that will allow for faster simulations of higher-order PDE models.  Promising candidates could be suitably adapted Lattice-Boltzmann algorithms~\cite{LB_Marenduzzo_hybrid} that implement negative~\lq viscosities\rq~\cite{1989Rothman}.

 \par
Finally, the theoretical analysis of higher-order PDEs poses a number of future mathematical challenges,
one of which being their  derivation from underlying microscopic, multi-component or kinetic models through systematic projection methods. 
{\color{black}{The good agreement of  the vector model with experimentally measured flow structures in quasi-2D~\textit{Bacillus subtilis} suspensions~\cite{2012Wensink} suggests 
that derivations of active continuum theories from microscopic models, which typically involve gradient expansions, should go beyond the most frequently considered second-order approximations.}
}
Another interesting question relates to the formulation of consistent boundary conditions at interfaces. This problem, which was circumvented in our simulations by considering toroidal domains, is also encountered in other higher-order structure formation and transport theories~\cite{2006Aranson,2009Cross,2011Bazant}.  To identify physically reasonable boundary conditions for active fluids at solid interfaces is not only relevant from a mathematical perspective but also from the experimental point of view, as it will directly affect predictions for effective shear viscosities \cite{Hess_BC} and other measurable quantities.
 {\color{black}{In particular, future simulations of the vector model in confined geometries can help to interpret  rheological measurements in microbial suspensions~\cite{2009SoAr,2011Aranson_PRE,2012Marenduzzo}, and they also promise insights into the effects~\cite{2012Woodhouse} of surface structures and microfluidic channel design on active fluids.}
}

\section*{Acknowledgements}
J. D. would like to thank Martin Bazant, Edgar Knobloch, Hartmut L\"owen, Cristina Marchetti, Ingmar Riedel-Kruse, Tim Pedley, Rik Wensink and Julia Yeomans for helpful discussions. This work was supported by the European Research Council (ERC), Advanced Investigator Grant 247333 (R.~E.~G. and J. D.) and by the Deutsche Forschungsgemeinschaft (DFG), GRK1558 (M. B. and S. H.).

\appendix

\section{Functional representation}

This appendix aims to summarize functional representations of the field equations for both the scalar model and the vector model. To this end, consider a functional $\mcal{F}$ that depends on some real-valued fields $\phi_k(x_1,\ldots,x_d), k=1,\ldots, N$, and their first and second derivatives, and can be written as
\be
\mathcal{F}[\mathbf \phi]=\int d^d x \,F(\phi_k, \partial_i \phi_k, \partial_{ij}\phi_k),
\ee
where  $\mathbf \phi=(\phi_k)$ and  $\partial_i=\partial/\partial x_i$, $\partial_{ij}=\partial^2/\partial x_i\partial x_j$.
Assuming $F(\eta_k,\xi_{ik},\zeta_{ijk})$ is a quadratic polynomial in $\xi_{ik}$ and $\zeta_{ijk}$, the functional derivative of   $\mcal{F}$ with respect to $\phi_k$ is given by
\be\label{a-e:func}
\f{\gd \mathcal{F}}{\gd \phi_k}
= 
\f{\partial F}{\partial \phi_k} -
\partial_i  \f{\partial F}{\partial (\partial_i \phi_k)} +
\partial_{ij}  \f{\partial F}{\partial (\partial_{ij} \phi_k)},
\ee
with a summation convention for identical indices $i,j=1,\ldots,d$.

\subsection{Swift-Hohenberg model}
\label{app:scalar}

Using \Eref{a-e:func} with $\mathbf \phi=\phi_1=\psi$,   the (pseudo-)scalar Swift-Hohenberg equation~\eref{e:scalar}
\begin{equation}\label{a-e:scalar}
\partial_t \psi = 
\gc_0 \triangle \psi  -\gc_2\triangle^2    \psi
-\f{\partial U}{\partial \psi} 
\end{equation}
can be written in the well-known form
\begin{equation}\label{a-e:scalar_2}
 \partial_t \psi   = -\f{\gd \mathcal{F}}{\gd \psi},
 \end{equation}
where
\be
\mathcal{F}[\psi]=\int d^d x\;\left[
\f{1}{2}\gc_0 (\nabla\psi)\cdot (\nabla\psi)  +\f{1}{2}\gc_2 (\triangle \psi )(\triangle \psi ) + U(\psi)
\right].
\ee

\subsection{Vector model}
\label{app:vector}

In component form, the vector model dynamics defined by Equations~\eref{e:div_free} and \eref{e:conti-c} reads
\begin{eqnarray}
\fl\qquad
\qquad\qquad\quad
0&=&\;\,\partial_i v_i,
\qquad\\
\fl\qquad
(\partial_t + \gl_0 v_i \partial_i) v_k
&=&
-\partial_k p  +\gl_1 \partial_k(v_j v_j) + 
\Gc_0 \partial_{ii} v_k -\Gc_2(\partial_{ii})^2    v_k
 - \f{\partial U}{\partial v_k}.
\label{a-e:conti-c}
\end{eqnarray}
Using \Eref{a-e:func} with $\mathbf \phi=(p,\mathbf v)$,  these field equations can be written as
\be
0=-\f{\gd \mathcal{F}}{\gd p}
,
\qquad\qquad
\left(\partial_t  +\gl_0 v_i  \partial_i\right) v_k
= -
\f{\gd \mathcal{F}}{\gd v_k},
\label{a-e:conti-2}
\ee
where
\be
\mathcal{F}[p,\mathbf v] 
&=&
\int d^d x\,\biggl[
p (\p_iv_i) -2\gl_1(v_k v_i \partial_i v_k ) +
\nonumber
\\
&&\qquad\quad\;
\f{1}{2}\Gc_0 (\partial_i v_k) (\partial_iv_k)  +
\f{1}{2}\Gc_2 (\partial_{ii} v_k )(\partial_{jj} v_k ) + 
U(\mathbf v)
\biggr].
\label{a-e:vec-free_energy}
\ee
As evident from the representation~\eref{a-e:conti-2}, apart from the convective derivative, which can be rescaled by active hydrodynamic stresses via $\gl_0$, see~\Eref{e:E}, the vector model dynamics can be understood in terms of an optimization of the  effective \lq free-energy\rq\space functional~\eref{a-e:vec-free_energy}.

\section*{References}

\providecommand{\newblock}{}



\begin{thebibliography}{10}
\expandafter\ifx\csname url\endcsname\relax
  \def\url#1{{\tt #1}}\fi
\expandafter\ifx\csname urlprefix\endcsname\relax\def\urlprefix{URL }\fi
\providecommand{\eprint}[2][]{\url{#2}}

\bibitem{2012Vicsek}
Vicsek T and Zafeiris A 2012 {\em Physics Reports\/} {\bf 517} 71--140

\bibitem{2010Ramaswamy}
Ramaswamy S 2010 {\em Annu. Rev. Cond. Mat. Phys.\/} {\bf 1} 323--345

\bibitem{2011KochSub}
Koch D~L and Subramanian G 2011 {\em Annu. Rev. Fluid. Mech.\/} {\bf 43}
  637--659

\bibitem{2009Parisi}
Cavagna A, Cimarelli A, Giardina I, Parisi G, Santagati R, Stefanini F and
  Viale M 2009 {\em Proc. Natl. Acad. Sci. USA\/} {\bf 107} 11865--11870

\bibitem{2011Couzin}
Katz Y, Ioannou C~C, Tunstro K, Huepe C and Couzin I~D 2011 {\em Proc. Natl.
  Acad. Sci. USA\/} {\bf 108} 18720--18725

\bibitem{1997Kessler}
Kessler J~O and Wojciechowski M 1997 {\em Collective Behavior and Dynamics of
  Swimming Bacteria\/} (Oxford, England: Oxford University Press) pp 417--450

\bibitem{2004DoEtAl}
Dombrowski C, Cisneros L, Chatkaew S, Goldstein R~E and Kessler J~O 2004 {\em
  Phys. Rev. Lett.\/} {\bf 93} 098103

\bibitem{2007Cisneros}
Cisneros L~H, Cortez R, Dombrowski C, Goldstein R~E and Kessler J~O 2007 {\em
  Exp. Fluids\/} {\bf 43} 737--753

\bibitem{2012Lin}
Liu K~A and I L 2012 {\em Phys. Rev. E\/} {\bf 86} 011924

\bibitem{2007SoEtAl}
Sokolov A, Aranson I~S, Kessler J~O and Goldstein R~E 2007 {\em Phys. Rev.
  Lett.\/} {\bf 98} 158102

\bibitem{Swinney_bactclust}
Zhang H~P, Be'er A, Florin E~L and Swinney H~L 2010 {\em Proc. Natl. Acad. Sci.
  USA\/} {\bf 107} 13626--13630

\bibitem{2012Peruani}
Peruani F, Starru{\ss} J, Jakovljevic V, S\/ogaard-Andersen L, Deutsch A and
  B\"ar M 2012 {\em Phys. Rev. Lett.\/} {\bf 108} 098102

\bibitem{2012Marchetti}
Marchetti M~C, Joanny J~F, Ramaswamy S, Liverpool T~B, Prost J, Rao M and
  {Aditi Simha} R Soft active matter arXiv:1207.2929v1

\bibitem{2012Wensink}
Wensink H~H, Dunkel J, Heidenreich S, Drescher K, Goldstein R~E, L\"{o}wen H
  and Yeomans J~M 2012 {\em Proc. Natl. Acad. Sci. USA\/} {\bf 109}
  14308--14313

\bibitem{2012Toner}
Toner J 2012 {\em Phys. Rev. E\/} {\bf 86} 031918

\bibitem{2012PawelReview}
Romanczuk P, B\"ar M, Ebeling W, Lindner B and Schimansky-Geier L 2012 {\em
  Eur. Phys. J. - Special Topics\/} {\bf 202} 1--162

\bibitem{2005ToTuRa}
Toner J, Tu Y and Ramaswamy S 2005 {\em Ann. Phys.\/} {\bf 318} 170--244

\bibitem{1998TonerTu_PRE}
Toner J and Tu Y 1998 {\em Phys. Rev. E\/} {\bf 58} 4828--4858

\bibitem{2008Wolgemuth}
Wolgemuth C~W 2008 {\em Biophys. J.\/} {\bf 95} 1564--1574

\bibitem{2009BaMa_PNAS}
Baskaran A and Marchetti M~C 2009 {\em Proc. Natl. Acad. Sci.\/} {\bf 106}
  15567--15572

\bibitem{LB_Marenduzzo_hybrid}
Marenduzzo D, Orlandini E, Cates M~E and Yeomans J~M 2007 {\em Phys. Rev. E\/}
  {\bf 76} 031921

\bibitem{2010Pedley}
Pedley T~J 2010 {\em Experimental Mechanics\/} {\bf 50} 1293--1301

\bibitem{2002Ra}
Simha R~A and Ramaswamy S 2002 {\em Phys. Rev. Lett.\/} {\bf 89} 058101

\bibitem{2008SaintillanShelley}
Saintillan D and Shelley M 2008 {\em Phys. Fluids\/} {\bf 20} 123304

\bibitem{2007Ar}
Aranson I~S, Sokolov A, Kessler J~O and Goldstein R~E 2007 {\em Phys. Rev. E\/}
  {\bf 75} 040901

\bibitem{2010Bausch}
Schaller V, Weber C, Semmrich C, Frey E and Bausch A~R 2010 {\em Nature\/} {\bf
  467} 73--77

\bibitem{2011Japan}
Ishikawa T, Yoshida N, Ueno H, Wiedeman M, Imai Y and Yamaguchi T 2011 {\em
  Phys. Rev. Lett.\/} {\bf 107} 028102

\bibitem{2012Lu}
Lu P~J, Giavazzi F, Angelini T~E, Zaccarelli E, Jargstorff F, Schofield A~B,
  Wilking J~N, Romanowsky M~B, Weitz D~A and Cerbino R 2012 {\em Phys. Rev.
  Lett.\/} {\bf 108} 218103

\bibitem{2012Sumino}
Sumino Y, Nagai K~H, Shitaka Y, Tanaka D, Yoshikawa K, Chate H and Oiwa K 2012
  {\em Nature\/} {\bf 483} 448--452

\bibitem{2005Riedel_Science}
Riedel I~H, Kruse K and Howard J 2005 {\em Science\/} {\bf 309} 300--303

\bibitem{2006Aranson}
Aranson I~S and Tsimring L~S 2006 {\em Rev. Mod. Phys.\/} {\bf 78} 641--692

\bibitem{2009Herminghaus}
Huang K, Roeller K and Herminghaus S 2009 {\em Eur. Phys. J. Special Topics\/}
  {\bf 179} 25--32

\bibitem{1990Knobloch}
Knobloch E 1990 {\em Physica D\/} {\bf 41} 450--479

\bibitem{1991Ray}
Goldstein R~E, Gunaratne G~H, Gil L and Coullet P 1991 {\em Phys. Rev. A\/}
  {\bf 43}(12) 6700--6721

\bibitem{2009Cross}
Cross M and Greenside H 2009 {\em Pattern Formation and Dynamics in
  Nonequilibrium Systems\/} (Cambridge, England: Cambridge University Press)

\bibitem{2010Nicoli}
Nicoli M, Vivo E and Cuerno R 2010 {\em Phys. Rev. E\/} {\bf 82} 045202(R)

\bibitem{1977SwiftHohenberg}
Swift J and Hohenberg P~C 1977 {\em Phys. Rev. A\/} {\bf 15} 319--328

\bibitem{2006Du}
Du Q, Liu C and Wang X 2006 {\em J. Comp. Phys.\/} {\bf 212}

\bibitem{1997Lifshitz}
Lifshitz R and Petrich D~M 1997 {\em Phys. Rev. Lett.\/} {\bf 79} 1261--1264

\bibitem{2011Bazant}
Bazant M~Z, Storey B~D and Kornyshev A~A 2011 {\em Phys. Rev. Lett.\/} {\bf
  106} 046102

\bibitem{2011Boyer}
Boyer D, Mather W, Mondrag\'on-Palomino O, Orozco-Fuentes S, Danino T, Hasty J
  and Tsimring L~S 2011 {\em Phys. Biol.\/} {\bf 8} 026008

\bibitem{2012Marenduzzo}
Foffano G, Lintuvuori J~S, Morozov A~N, Stratford K, Cates M~E and Marenduzzo D
  2012 {\em Eur. Phys. J. E\/} {\bf 35} 98

\bibitem{2012Woodhouse}
Woodhouse F and Goldstein R~E 2012 {\em Phys. Rev. Lett.\/} {\bf 109} 168105

\bibitem{2009Rafai}
Rafai S, Jibuti L and Peyla P 2009 {\em Phys. Rev. Lett.\/} {\bf 104} 098102

\bibitem{2009SoAr}
Sokolov A and Aranson I~S 2009 {\em Phys. Rev. Lett.\/} {\bf 103} 148101

\bibitem{2011Aranson_PRE}
Ryan S~D, Haines B~M, Beryland L, Ziebert F and Aranson I~S 2011 {\em Phys.
  Rev. E\/} {\bf 83} 050904(R)

\bibitem{2010Austin_PRL}
Lambert G, Liao D and Austin R~H 2010 {\em Phys. Rev. Lett.\/} {\bf 104} 168102

\bibitem{2012Kantsler_PNAS}
Kantsler V, Dunkel J, Polin M and Goldstein R~E 2012 {\em Proc. Natl. Acad.
  Sci. USA\/}  in press

\bibitem{2008BaMa}
Baskaran A and Marchetti M~C 2008 {\em Phys. Rev. E\/} {\bf 77} 011920

\bibitem{2009Bertin}
Bertin E, Droz M and Gr\'egoire G 2009 {\em J. Phys. A: Math. Theor.\/} {\bf
  42} 445001

\bibitem{2012Peshkov_PRL}
Peshkov A, Aronson I~S, Bertin E, Chat\'e H and Ginelli F 2012 {\em Phys. Rev.
  Lett.\/}  in press

\bibitem{1981Volkenstein}
Belintsev B~N, Livshitz M~A and Volkenstein M~V 1981 {\em Physics Letters\/}
  {\bf 82A} 375--377

\bibitem{1981Volkenstein_2}
Belintsev B~N, Livshitz M~A and Volkenstein M~V 1981 {\em Z. Phys. B -
  Condensed Matter\/} {\bf 44} 345--351

\bibitem{2007Hutt}
Hutt A 2007 {\em Phys. Rev. E\/} {\bf 75} 026214

\bibitem{2005Berg_Nature}
DiLuzio W~R, Turner L, Mayer M, Garstecki P, Weibel D~B, Berg H~C and
  Whitesides G~M 2005 {\em Nature\/} {\bf 435} 1271--1274

\bibitem{2012Grossmann}
Gro{\ss}mann R, Schimansky-Geier L and Romanczuk P 2012 {\em New J. Phys.\/}
  {\bf 14} 073033

\bibitem{Orszag}
Gottlieb D and Orszag S~A 1977 {\em Numerical analysis of spectral methods:
  Theory and applications\/} (Montpelier, Vermont, USA: SIAM)

\bibitem{Pedrosa}
Pedrosa J, Hoyuelos M and Martel C 2008 {\em Eur. Phys. J. B\/} {\bf 66}
  525--530

\bibitem{Canuto}
Canuto C, Hussaini M~Y, Quarteroni A and Zang T~A 2006 {\em Spectral methods:
  fundamentals in single domains\/} (Berlin, Germany: Springer)

\bibitem{1997Swinney}
Petrov V, Ouyang Q and Swinney H~L 1997 {\em Nature\/} {\bf 388} 655--657

\bibitem{2006DuHi}
Dunkel J and Hilbert S 2006 {\em Physica A\/} {\bf 370} 390--406

\bibitem{2006HiDu}
Hilbert S and Dunkel J 2006 {\em Phys. Rev. E\/} {\bf 74} 011120

\bibitem{2008Kastner}
Kastner M 2008 {\em Rev. Mod. Phys.\/} {\bf 80}(1) 167--187

\bibitem{2008Tang_PNAS}
Li G, Tam L~K and Tang J~X 2008 {\em Proc. Natl. Acad. Sci. USA\/} {\bf 105}
  18355--18359

\bibitem{2010Elgeti_BiophysJ}
Elgeti J, Kaupp U~B and Gompper G 2010 {\em Biophys. J.\/}  1018--1026

\bibitem{2012Dunstan}
Dunstan J, Mino G, Clement E and Soto R 2012 {\em Phys. Fluids\/} {\bf 24}
  011901

\bibitem{1980Gibbins_Circle}
Gibbins B~H 1980 {\em J. Cell Biol.\/} {\bf 84} 1--12

\bibitem{2003Bazant}
Hadjiconstantinou N~J, Garcia A~L, Bazant M~Z and He G 2003 {\em J. Comp.
  Phys.\/} {\bf 187} 274--297

\bibitem{2011Cisneros_PRE}
Cisneros L~H, Kessler J~O, Ganguly S and Goldstein R~E 2011 {\em Phys. Rev.
  E\/} {\bf 83} 061907

\bibitem{2011DrescherEtAl}
Drescher K, Dunkel J, Cisneros L~H, Ganguly S and Goldstein R~E 2011 {\em Proc.
  Natl. Acad. Sci. USA\/} {\bf 108} 10940--10945

\bibitem{2010DrEtAl_PRL}
Drescher K, Goldstein R~E, Michel N, Polin M and Tuval I 2010 {\em Phys. Rev.
  Lett.\/} {\bf 105} 168101

\bibitem{2004Frisch}
Frisch U 2004 {\em Turbulence\/} (Cambridge, England: Cambridge University
  Press)

\bibitem{2010Mishra_JSM}
Mishra S, Simha R~A and Ramaswamy S 2010 {\em J. Stat. Mech.: Theor. Exp.\/}
  P02003

\bibitem{1989Rothman}
Rothman D~H 1989 {\em J. Stat. Phys.\/} {\bf 56} 517--524

\bibitem{Hess_BC}
Heidenreich S, Ilg P and Hess S 2007 {\em Phys. Rev. E\/} {\bf 75} 066302

\end{thebibliography}
\end{document}